\documentclass[%
reprint, 
nofootinbib,
 amsmath,amssymb,
 aps,
]{revtex4-1}
\usepackage{epsfig,epsf,rotating,wrapfig,tabularx}
\usepackage{dcolumn}% Align table columns on decimal point
\usepackage{bm}% bold math
\usepackage{graphicx}% Include figure files
\begin{document}

\preprint{APS/123-QED}

\title{A photon-proton marriage. }% Force line breaks with \\
%\thanks{A footnote to the article title}%

\author{\firstname{A.~A.}~\surname{Bylinkin}}
 \email{alexander.bylinkin@desy.de}
\affiliation{%
 Institute for Theoretical and Experimental
Physics, ITEP, Moscow, Russia
}%
% \altaffiliation[Also at ]{Physics Department, XYZ University.}%Lines break automatically or can be forced with \\
\author{\firstname{A.~A.}~\surname{Rostovtsev}}
 \email{rostov@itep.ru}
\affiliation{%
 Institute for Theoretical and Experimental
Physics, ITEP, Moscow, Russia
}%

%\date{\today}% It is always \today, today,
             %  but any date may be explicitly specified

\begin{abstract}
%% Text of abstract
The shapes of invariant differential cross sections for charged hadron production as function of hadrons transverse momentum and rapidity in $ep$ collisions at HERA machine are considered. The particle spectra shapes observed in $pp$ and $\gamma\gamma$ collisions before have shown very different properties. This difference is explained in terms of the introduced qualitative model for hadroproduction. Moreover, it could be directly measured in the “mixed” type collisions of photon and proton at HERA experiments. Finally, such measurement that can be performed at HERA is proposed.
\end{abstract}

\pacs{Valid PACS appear here}% PACS, the Physics and Astronomy
                             % Classification Scheme.
%\keywords{Suggested keywords}%Use showkeys class option if keyword
                              %display desired
\maketitle

%\tableofcontents

\section{Introduction}
\label{part1}
Inclusive charged particle distributions have been studied for a long time to derive the general properties of hadronic interactions at high energies. A large body of the experimental data on charge particle production spectra in baryon-baryon, gamma-baryon and gamma-gamma collisions has been accumulated during past decades. However, the underlying dynamics of hadron production in high energy particle interactions is still not fully understood. A comparative detailed analysis of the measured spectra of charged particles produced in different type of collisions could shed light on the hadroproduction mechanisms.   
\par
A new unified approach, which was shown to provide a better description of charged particle production spectra shape then the one traditionally used~\cite{Tsallis}, was proposed recently~\cite{OUR1}. It was suggested to approximate the charged particle spectra as function of the particle’s transverse momentum by a sum of an exponential (Boltzmann-like) and a power law statistical distributions:
\begin{equation}
\label{eq:exppl}
\frac{d\sigma}{P_T d P_T} = A_e\exp {(-E_{Tkin}/T_e)} +
\frac{A}{(1+\frac{P_T^2}{T^{2}\cdot n})^n},
\end{equation}
where  $E_{Tkin} = \sqrt{P_T^2 + M^2} - M$
with M equal to the produced hadron mass. $A_e, A, T_e, T, n$ are the free parameters to be determined by fit to the data.  The detailed arguments for this particular choice are given in~\cite{OUR1}.  For the charged hadron spectra a mass of hadrons is assumed to be equal to the pion mass. 

\begin{figure}[h]
\includegraphics[width =8cm]{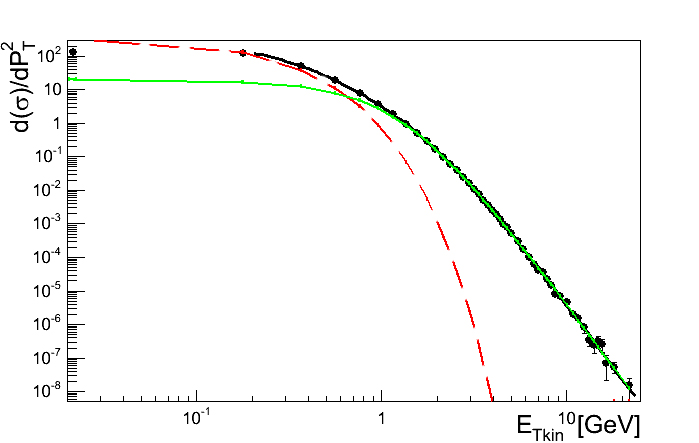}
\caption{\label{fig:0} Charge particle differential cross section~\cite{UA1} fitted to the function~(\ref{eq:exppl}): the red (dashed) line shows the exponential term and the green (solid) one - the power law.}
\end{figure}

A typical charged particle spectrum as function of transverse energy, fitted with this function~(\ref{eq:exppl}) is shown in Fig~\ref{fig:0}. 
The contributions of the exponential and power law
terms
of the parameterization~(\ref{eq:exppl}) to
the typical spectrum of charged particles produced in $pp-$collisions are also shown separately in Figure~\ref{fig:0}. As one can see the exponential term dominates the particle spectrum at low $P_T$ values.

The relative contribution of these terms is characterized by ratio $R$ of the power law term alone to the parameterization function integrated over $P_T^{2}$:

\begin{equation}
R = \frac{AnT}{AnT + A_e(2MT_e + 2T_e^2)(n-1)}
\end{equation}

It was found ~\cite{OUR1} that the exponential contribution dominates the charged particle spectra in $pp$ collisions~\cite{UA1} while it is completely missing in $\gamma\gamma$ interactions~\cite{OPAL}. Moreover, the exponential contribution is characteristic for charged pion production and is much less pronounced in kaon or proton (antiproton) production spectra~\cite{OUR2}. 

\section{Qualitative model}
\label{part2}
\begin{figure*}[!ht]
\includegraphics[width =8cm]{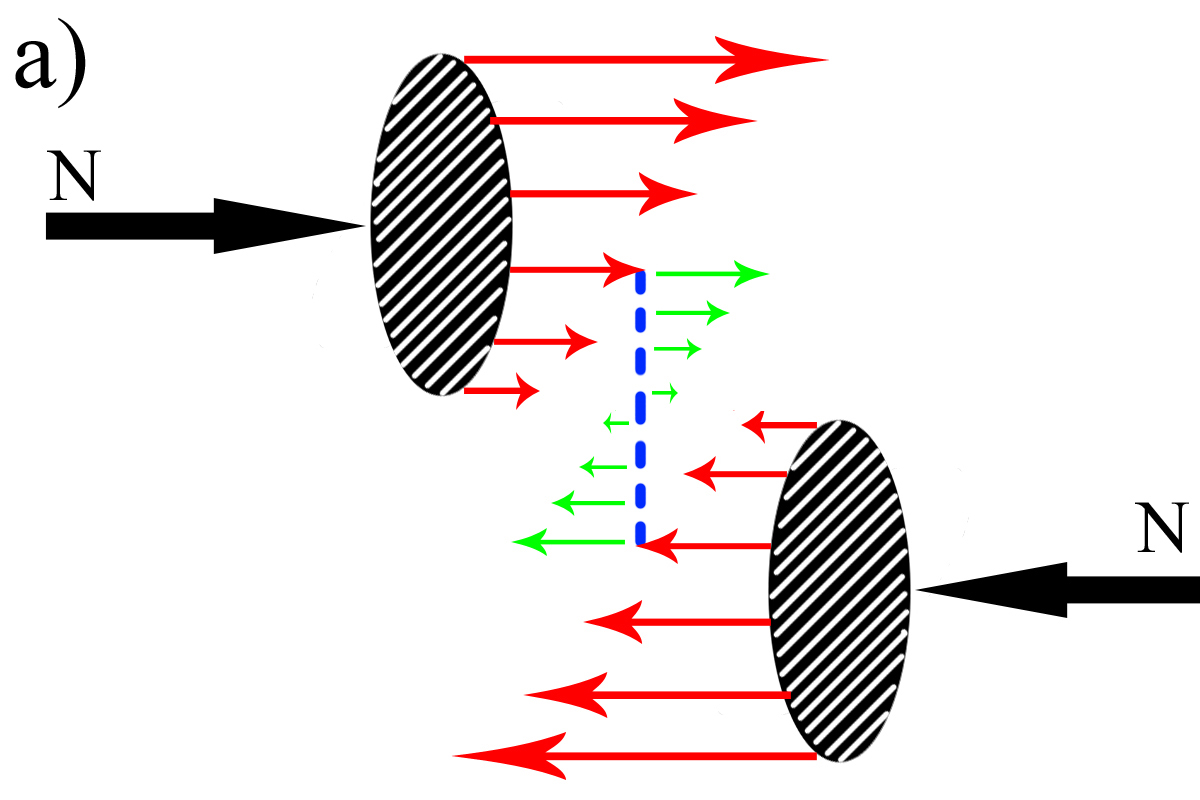}
%\caption{\label{fig} Two different sources of hadroproduction: red arrows - particles produced by the preexisted partons, green - particles produced via the Pomeron exchange.}
%\end{figure}
\quad
%\begin{figure}[h]
\includegraphics[width =8cm]{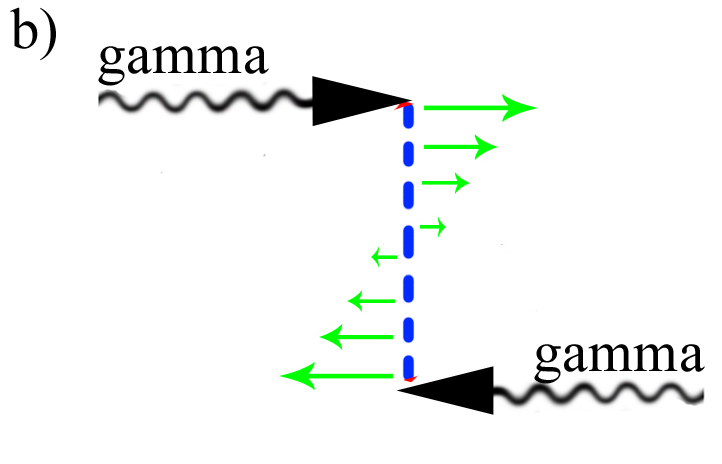}
\caption{\label{fig} Two different sources of hadroproduction in high energy baryonic (a) and $\gamma\gamma$ (b) collisions: red arrows - particles produced by the preexisted partons, green - particles produced via the Pomeron exchange.}
\end{figure*}

The hadroproduction process in baryon-baryon high energy interactions could be decomposed into at least two distinct parts. These parts are characterized by two different sources of produced hadrons. The first one is associated with the baryon valence quarks and a quark-gluon cloud coupled to the valence quarks. Those partons preexist long time before the interaction and could be considered as being a thermalized statistical ensemble. When a coherence of these partonic systems is destroyed via strong interaction between the two colliding baryons these partons hadronize into particles released from the collision. The hadrons from this source are distributed presumably according to the Boltzmann-like exponential statistical distribution in transverse plane w.r.t. the interaction axis~\cite{Hagedorn:1965st}.
\par
The second source of hadrons is directly related to the virtual partons exchanged between two colliding partonic systems. In QCD this mechanism is described by the BFKL Pomeron exchange~\cite{Fadin:1998py}. The radiated partons from this Pomeron have presumably a typical for the pQCD power law spectrum~\cite{Eskola:1996mb}. Schematically Figure~\ref{fig}  shows these two sources of particles produced in high energy baryonic (a) and $\gamma\gamma$ (b)  collisions. Obviously, there are no thermalized partons preexisted long time before the $\gamma\gamma$ interactions. This simplified model explains the observed absence of the Boltzmann-like term in the spectra of hadrons produced in $\gamma\gamma$ collisions. This explanation is qualitative, however. 

\section{$ep$ collisions at HERA}
\label{part3}
\begin{figure}[h]
\includegraphics[width =8cm]{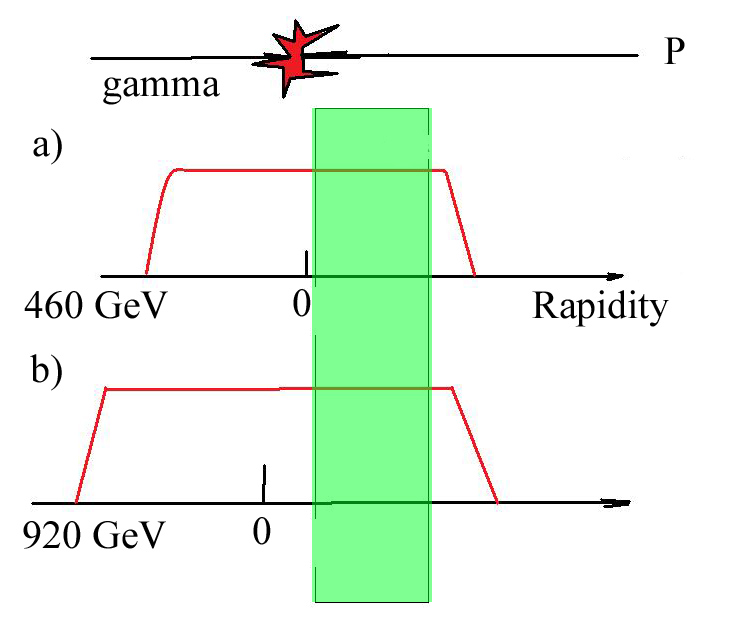}
\caption{\label{fig2} Charge particle rapidity distributions at HERA for reduced ($E_p = 460$ GeV) (a) and nominal ($E_p = 920$ GeV) (b) beam energies. Green area shows a typical coverage in rapidity space of the central tracking detectors at HERA experiments.}
\end{figure}

Previously it was found~\cite{OUR1} that charge particle transverse momentum distributions measured at $ep$-collider HERA do not show any sizable contribution of the exponential Boltzmann-like part of the spectrum similar to that found in $\gamma\gamma$ collisions. At HERA both photoproduction and DIS processes are mediated by exchange photon. Due to the strong asymmetry in the energies of colliding photon and proton at HERA these inclusive measurements~\cite{H1_0, H1_1} were made entirely on the photon side of the event rapidity space. Figure ~\ref{fig2} shows a typical coverage in rapidity space of the central tracking detectors at HERA experiments.

\begin{figure}[h]
\includegraphics[width =8cm]{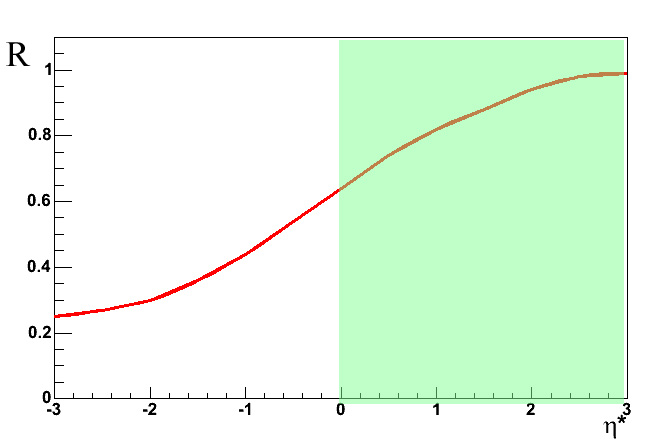}
\caption{\label{fig3} A qualitative picture of the expected power law term contribution of ~(\ref{eq:exppl}) in charge particle spectra as function of pseudorapidity ($\eta^*$). The green area display a rapidity region available for high precision track measurement by HERA experiments detectors for runs with reduced proton beam energy.}
\end{figure}

In  the proposed hadroproduction model it is natural to expect that chagred particle spectra produced in the photon hemisphere of the event resulting from $\gamma p$ interaction at HERA carry similarities to that in $\gamma\gamma$ collisions. At the same time the proton hemisphere of the event resulting from $\gamma p$ interaction at HERA carries similarity to $pp$ collisions.  This implies that the composition of the hadron spectra as measured on the photon and proton sides at HERA have a large quantitative difference in the Boltzmann term contribution to these spectra.  Unfortunately, the proton hemisphere of the gamma-proton collision events at HERA is not reachable for accurate track measurements both in H1 and ZEUS detectors. However, it is likely to observe a transition region in the rapidity space around event central rapidity values where one could observe a smooth transition between two different types of the spectra shapes: one is characteristic to the photon and another to baryon regimes described above. Qualitatively this expected behavior is illustrated in Fig ~(\ref{fig3}). The experimental challenge for such measurement is to find a way for an accurate and reliable track measurements close to the zero rapidity values in the $\gamma p$ center of mass system.  Figure ~(\ref{fig2}) shows, that the maximal reach in the rapidity space towards the proton hemisphere at HERA could be obtained using a set of the data collected with reduced proton beam energy. At the same time one should keep a selection of the $ep$ events with high energy of the photon mediated these interactions. Such reduced energy data were specially collected in $2007$ at HERA in order to facilitate high precision $F_L$ proton structure function measurement~\cite{FL}.
\section{Conclusion}
\label{part4}
In conclusion, we propose to measure the charged particle spectra produced in $ep$ collision at HERA as function of pseudorapidity of produced particles in $\gamma p$ center of mass system and to analyze them in terms of the introduced qualitative model. This measurement has to be made within whole rapidity space available for the measurement including the central rapidities as much as possible close to the proton hemisphere. In order to make this measurement more conclusive it is to be made using the reduced proton energy special data sets at best.

%\end{acknowledgments}
%% The Appendices part is started with the command \appendix;
%% appendix sections are then done as normal sections
%% \appendix

%% \section{}
%% \label{}

%% References
%%
%% Following citation commands can be used in the body text:
%% Usage of \cite is as follows:
%%   \cite{key}          ==>>  [#]
%%   \cite[chap. 2]{key} ==>>  [#, chap. 2]
%%   \citet{key}         ==>>  Author [#]

%% References with bibTeX database:

\bibliographystyle{model1a-num-names}
%\bibliography{<your-bib-database>}

%% Authors are advised to submit their bibtex database files. They are
%% requested to list a bibtex style file in the manuscript if they do
%% not want to use model1a-num-names.bst.

%% References without bibTeX database:

% \begin{thebibliography}{00}

%% \bibitem must have the following form:
%%   \bibitem{key}...
%%

% \bibitem{}

% \end{thebibliography}

\end{document}